\documentclass[hyper]{JHEP} 

\usepackage{epsfig}

\newcommand\fverb{\setbox\pippobox=\hbox\bgroup\verb}

\newcommand\fverbdo{\egroup\medskip\noindent%

            \fbox{\unhbox\pippobox}\ }

\newcommand\fverbit{\egroup\item[\fbox{\unhbox\pippobox}]}

\newbox\pippobox

\title{Note About Redefinition of BRST Operator for
Pure Spinor String in General
Background}
\author{by J. Kluso\v{n}\\
     Department of Theoretical Physics and Astrophysics\\
                   Faculty of Science, Masaryk University\\
Kotl\'{a}\v{r}sk\'{a} 2, 611 37, Brno\\
Czech Republic\\
    E-mail: \email{klu@physics.muni.cz}}
\preprint{ \\
\hepth{0803.4390}} \abstract{We discuss
an analysis presented by N. Berkovits
in [arXiv:0712.0324] in the context of
classical mechanics and perform its
extension  to  the case of pure spinor
string in general background.}
 \keywords{string theory}

\newcommand{\cP}{\mathcal{P}}

\def\pb  #1{\left\{#1\right\}}

\newcommand{\hp}{\hat{p}}

\newcommand{\hJ}{\hat{J}}

\newcommand{\mH}{\mathcal{H}}

\newcommand{\bH}{\mathbf{H}}

\newcommand{\htheta}{\hat{\theta}}

\newcommand{\hd}{\hat{d}}
\newcommand{\halpha}{\hat{\alpha}}
\newcommand{\hbeta}{\hat{\beta}}
\newcommand{\hdelta}{\hat{\delta}}
\newcommand{\hgamma}{\hat{\gamma}}
\newcommand{\hlambda}{\hat{\lambda}}

\newcommand{\hw}{\hat{w}}
\newcommand{\hN}{\hat{N}}

\newcommand{\hmu}{\hat{\mu}}

\newcommand{\mL}{\mathcal{L}}

\newcommand{\tP}{\tilde{\cP}}
\newcommand{\hP}{\hat{P}}
\newcommand{\hpi}{\hat{\pi}}
\newcommand{\hOmega}{\hat{\Omega}}

\newcommand{\hC}{\hat{C}}

\newcommand{\homega}{\hat{\omega}}

\newcommand{\tgamma}{\tilde{\gamma}}
\newcommand{\tbeta}{\tilde{\beta}}
\newcommand{\bG}{\mathbf{G}}
\begin{document}
\section{Introduction}
The pure spinor formalism
 is a super-Poincare covariant description
of the super-string
\cite{Berkovits:2000fe} \footnote{For
review, see
\cite{Berkovits:2002zk,Grassi:2005av,Grassi:2003cm,Grassi:2002sr,Nekrasov:2005wg}.}.
This new formulation has many
attractive properties, for example, it
simplifies calculation of multiloop
amplitudes
\cite{Berkovits:2004px,Berkovits:2006vi,Mafra:2008ar}.
Further it allows to find quantum
formulation of superstring in the
background with Ramond-Ramond
background, at least in principle
\cite{Berkovits:2000yr,Berkovits:2002zv,Berkovits:2004xu}
\footnote{For some related works, see
\cite{Mikhailov:2007eg,Mikhailov:2007mr,Berkovits:2007rj,
Berkovits:2007zk,Grassi:2006tj,Puletti:2006vb,Bianchi:2006im,
Kluson:2006wq,Berkovits:2004jw,Vallilo:2003nx,Vallilo:2002mh}.}.

On the other hand due to the fact that
the  BRST operator in the pure spinor
formalism is unconventional
 the relation of this
formalism to the Green-Schwarz (GS) and
Ramond-Neveu-Schwarz (RNS) formalisms
for the super-string was mysterious
\footnote{The problem how pure spinor
formalism arises from the conventional
GS formalism was attached in many
papers from several point of view
\cite{Berkovits:2001us,Oda:2001zm,Grassi:2004we,
Guttenberg:2004ht,Grassi:2003kq,Grassi:2003kq,
Oda:2005sd,Aisaka:2005vn,Chesterman:2004xt,
Aisaka:2003mw,Matone:2002ft,Grassi:2002xf,
Berkovits:2005bt,Berkovits:2004tw,Gaona:2005yw,Hoogeveen:2007tu}.}.
In a recent remarkable paper N.
Berkovits \cite{Berkovits:2007wz}
explained these features. His idea was
to add
 a pair of non-minimal fields to the theory
  and perform a similarity transformation
such that the pure spinor BRST operator
is expressed as a conventional-looking
BRST operator that contains collection
of first-class constraints. More
precisely this conventional-looking
BRST operator involves the Virasoro
constraints and twelve fermionic
constraints, where eleven of these
fermionic constraints are associated to
the eleven independent components of
the original bosonic pure spinor ghost.
The additional  fermionic constraints
and the Virasoro constraints are
associated to the new pair of
non-minimal fields, bosonic
($\tbeta,\tgamma$) and fermionic ($b,
c$). Even if  this conventional form of
the BRST operator is not manifestly
Lorentz invariant, it was shown that it
is useful for construction of
 GSO(-) vertex operators
and for relating the pure spinor
formalism to the GS and RNS formalisms.

Since the analysis presented in
\cite{Berkovits:2007wz} was very
interesting  the goal of this paper is
to apply the similar procedure for the
pure spinor string in general
background. Explicitly, our starting
point is the pure spinor action in
general background that was introduced
 in \cite{Berkovits:2001ue} \footnote{For
discussion of pure spinor string in
general background, see
\cite{Benichou:2008it,Chandia:2007vp,Adam:2007ws,
Chandia:2006ix,Chandia:2003hn,Berkovits:2002uc}.}.
As opposite to the original work
\cite{Berkovits:2001ue} we formulate
the pure spinor action  with general
world-sheet metric however keeping in
mind an important point that pure
spinor string is defined on the
world-sheet with flat metric. An
advantage of this formulation (That of course
should be considered as technical tool)
is that we can  easily find
Hamiltonian density
as a combination of Virasoro
constraints that play a prominent role
in Berkovits construction.
Then we  perform the
similarity transformation as in the
case of pure spinor string in flat
space-time. Since we consider the
background as general as possible we
do not try to calculate Poisson
brackets between $T_\pm$ explicitly.
Our basic presumption is
 that
 the Poisson brackets
between $T_\pm$ take standard form.
 Then we  argue
for an existence of two ghost number
$-1$ functions $G_\pm$ that play the
role of  $b_\pm$ ghost fields in
the standard formulation. We
analyze their Poisson brackets among
themselves and with Virasoro
constraints. Then we construct
 operator $R$ and we discuss its
 basic properties. We show that
 generally
this operator is time dependent and
we discuss  consequence
of its time dependence on the form
of the new BRST operator $Q'$.

The organization of this paper is as
follows. In the next section
(\ref{second}) we review the classical
treatment of the pure spinor string in
flat Minkowski background. We introduce
 basic notations and conventions. We
also review the approach presented in
\cite{Berkovits:2007wz} now formulated
in the context of classical mechanics
and Poisson brackets.

Then in section (\ref{third}) we
generalize this analysis to the case of
pure spinor string in general
background. In the first step we
develop Hamiltonian formalism for this
pure spinor string and find the form of
corresponding Hamiltonian and BRST
charges. Then we  argue for an
existence of two functions $G_\pm$ that
allow to express $T_\pm$ as a result of
the Poisson brackets of $Q$ with
$G_\pm$. Since we consider pure spinor
string in general background we will
not be able to find explicit form of
$G_\pm$ and their Poisson brackets with
$T_\pm$. On the other hand we can guess
the form of these Poisson brackets and
try to  analyze a consequence of these
non-trivial Poisson brackets on the
form of the operator $R$. Then we
determine the  new BRST operator $Q'$
that now contains collection of
first-class constraints. Namely, if we
take pure spinor constraints into
account then the new BRST operator $Q'$
contains collection of two Virasoro
constraints and $22$ fermionic
constraints. Finally, in conclusion
(\ref{fourth}) we outline our results
and suggest possible extension of this
work.
\section{Redefinition of Pure Spinor
String BRST Charge in Flat Background}\label{second}
In this section we review the
approach presented in \cite{Berkovits:2007wz}.
We perform this analysis in the
context of classical Hamiltonian dynamics
in order to have a contact with calculation
presented in next section.

Our starting point is pure
spinor string action in flat background
\begin{eqnarray}\label{Spureflat}
S&=&-\int
d^2\sigma\sqrt{-h}(\frac{1}{2}h^{\mu\nu}
\partial_\mu x^m\partial_\nu x^n
\eta_{mn}+\omega_{\mu \alpha}
\cP^{\mu\nu}\partial_\nu\theta^\alpha
+\homega_{\mu\halpha}\tP^{\mu\nu}
\partial_\nu\htheta^{\halpha}+
\nonumber \\
&+&w_{\mu \alpha}\cP^{\mu\nu}
\partial_\nu\lambda^\alpha
+\hw_{\mu\halpha}\tP^{\mu\nu}
\partial_\nu \hlambda^{\halpha}) \ ,
\nonumber \\
\end{eqnarray}
where $h^{\mu\nu}$ is two dimensional
world sheet metric, $\sigma^0=\tau \ ,
\sigma^1=\sigma$. Further,
$\alpha=1,\dots ,16$ label
Majorana-Weyl spinors and
$\halpha=1,\dots,16$ label second
Majorana-Weyl spinors and
$\omega_{\mu\alpha},
\homega_{\mu\halpha}$ are related to
the momenta conjugate to
$\theta^\alpha, \htheta^{\halpha}$.
$\gamma^m_{\alpha\beta} \ ,
\gamma^m_{\halpha\hbeta}$ are $16\times
16$ symmetric Dirac matrices. We also
introduced  chiral and (anti-chiral)
operators
\begin{equation}\label{projectordef}
\cP^{\mu\nu}=h^{\mu\nu}-
\frac{\varepsilon^{\mu\nu}}{\sqrt{-h}}
\ , \quad
\tP^{\mu\nu}=h^{\mu\nu}+\frac{\varepsilon^{\mu\nu}}{\sqrt{-h}}
\ ,
\end{equation}
where
$\varepsilon^{\tau\sigma}=-\varepsilon^{\sigma\tau}=1$.
Our goal is to develop classical
Hamiltonian formalism for pure spinor
action (\ref{Spureflat}).
From (\ref{Spureflat})
we determine momenta conjugate
to $x^m,\theta^\alpha,\htheta^{\halpha}$
\begin{eqnarray}
p_m &=&\frac{\delta S}{\delta
\partial_\tau x^m}=
-\sqrt{-h}h^{\tau\mu}\partial_\mu x^m \
, \nonumber \\
p_\alpha &=&\frac{\delta
S}{\delta\partial_\tau \theta^\alpha}=
\sqrt{-h}\omega_{\mu\alpha}\cP^{\mu\tau}
 \ , \quad
 p_{\halpha}=\frac{\delta S}{\delta
 \partial_\tau \htheta^{\halpha}}=
 \sqrt{-h}\homega_{\mu\halpha}\tP^{\mu\tau}
 \ .
 \nonumber \\
 \end{eqnarray}
In the same way we  proceed in case
of pure spinors and we define momenta
$\pi_\alpha,\hpi_{\halpha}$ conjugate
to $\lambda^\alpha,\hlambda^{\halpha}$
as
\begin{eqnarray}
\pi_{\alpha}=\frac{\delta S}{\delta
\partial_\tau \lambda^\alpha}=
-\sqrt{-h}w_{\mu\alpha}\cP^{\mu \tau} \
, \quad
\hpi_{\halpha}=\frac{\delta S}{\delta
\partial_\tau \hlambda^{\halpha}}=
-\sqrt{-h}\hw_{\mu\halpha}\tP^{\mu
\tau} \ .
\nonumber \\
\end{eqnarray}
However in case of pure
spinors there is slight
subtlety due to the pure
spinor constraints:
\begin{equation}
\lambda^\alpha (\gamma^m)_{\alpha\beta}
\lambda^\beta=0 \ , \quad
\hlambda^{\halpha}
(\gamma^m)_{\halpha\hbeta}
\hlambda^{\hbeta}=0 \ .
\end{equation}
These relations imply that not all
$\lambda$'s are independent
\footnote{We can find set of eleven
independent variables when we solve
these constraints in $U(5)$ invariant
manner. Under $SU(5)\times U(1)$, an
$SO(10)$ spinor decomposes as
$\lambda^\alpha \rightarrow
(\lambda^+,\lambda_{ab},\lambda^b)$
where $a=1$ to $5$,
$\lambda_{ab}=-\lambda_{ba}$ and
$(\lambda^+,\lambda_{ab},\lambda^a)$
carries $U(1)$ charge $(\frac{5}{2},
\frac{1}{2},-\frac{3}{2})$. If
$\lambda^+$ is assumed to be nonzero,
$\lambda \gamma^m\lambda$ implies that
\begin{equation}
\lambda^a=-\frac{1}{8\lambda^+}
\epsilon^{abcde}\lambda_{bc}\lambda_{de}
\
\end{equation}
so that $\lambda^\alpha$ has eleven
independent components parameterized by
$\lambda^+$ and $\lambda_{ab}$.}. On
the other hand in case of classical
calculations presented in this paper we
do not have to worry about pure spinor
constraints and in all calculations we
can tread all $\pi$'s and $\lambda$'s
as independent. Only in the end of the
calculations when we  count number of
independent constraints we use the
explicit parameterization of pure
spinors given in footnote.

Let us now return to the review of
basic properties of canonical
variables. By definition they obey
graded Poisson brackets
\begin{eqnarray}\label{cpbf}
\pb{x^m(\sigma),p_n(\sigma')}&=&
\delta^m_n\delta(\sigma-\sigma') \ , \nonumber \\
\pb{\theta^\alpha(\sigma),p_{\beta}(\sigma')}&=&
-\delta_\beta^\alpha\delta(\sigma-\sigma')
\ , \quad
\pb{\htheta^{\halpha}(\sigma),\hp_{\hbeta}(\sigma')}=
-\delta_{\hbeta}^{\halpha}\delta(\sigma-\sigma')
\ ,
 \nonumber \\
\pb{\lambda^\alpha(\sigma),\pi_\beta(\sigma')}&=&
\delta_\beta^\alpha
\delta(\sigma-\sigma') \ , \quad
\pb{\hlambda^{\halpha}(\sigma),\hpi_{\hbeta}(\sigma')}=
\delta_{\hbeta}^{\halpha}
\delta(\sigma-\sigma') \ .
\nonumber \\
\end{eqnarray}

Further  the pure spinor action has to
be accompanied with the BRST operators
$Q=Q_L+Q_R$ where two BRST charges take
the form
\begin{equation}\label{QLR}
Q_L=\int d\sigma \lambda^\alpha
d_\alpha \ , \quad  Q_R=\int d\sigma
\hlambda^{\halpha} \hd_{\halpha} \ ,
\end{equation}
where
\begin{eqnarray}
d_\alpha &=&
p_\alpha-ip_m(\gamma^m\theta)_\alpha
+(\gamma^n\theta)_\alpha
(\theta\gamma^m\partial_\sigma\theta)\eta_{mn}
+i(\gamma^m\theta)_\alpha
\partial_\sigma x^n\eta_{mn} \ ,
\nonumber \\
\hd_{\halpha}&=&\hp_{\halpha} -ip_m
(\gamma^m \htheta)_{\halpha}-
(\gamma^m\htheta)_{\halpha}(\htheta\gamma^n
\partial_\sigma\theta)\eta_{mn}
-i(\gamma^m\htheta)_{\halpha}\partial_\sigma
x^n\eta_{mn}\ .
\nonumber \\
\end{eqnarray}

We would like to stress that even if we
formulated  the action
(\ref{Spureflat}) with the general
world-sheet metric $h_{\mu\nu}$ the
pure spinor string theory is formulated
on the world-sheet with  flat
world-sheet metric where
$h_{\mu\nu}=\eta_{\mu\nu}\equiv
\mathrm{diag}(-1,1)$ \footnote{Pure
spinor string on world-sheet with
general metric was also studied in
interesting paper
\cite{Hoogeveen:2007tu}.}. Reason why
we consider theory with general
world-sheet metric is that we can
easily develop the Hamiltonian
formalism and also find the form of the
Virasoro constraints. Explicitly,  let
us introduce variables
\begin{equation}
\rho^\pm=\frac{\sqrt{-h} \pm
h_{\tau\sigma}}{h_{\sigma\sigma}} \ ,
\quad
\xi=\ln h_{\sigma\sigma} \ ,
\end{equation}
where $\rho_\pm$ are manifestly
invariant under Weyl transformation
$h'_{\mu\nu}= e^\phi h_{\mu\nu}$ while
$\xi$ transform as $\xi'=\xi+\phi$.
Using this notation we can express
the projectors (\ref{projectordef}) as
\begin{eqnarray}\label{projectors}
\cP^{\tau\tau}&=&-4\frac{e^{-\xi}}{(\rho^++\rho^-)^2}
\ , \quad
\cP^{\sigma\sigma}=4\frac{\rho^+\rho^-e^{-\xi}}
{(\rho^++\rho^-)^2}  \ , \nonumber \\
\cP^{\tau\sigma}&=&-4\frac{\rho^-e^{-\xi}}
{(\rho^++\rho^-)^2}  \ , \quad
\cP^{\sigma\tau}=4\frac{\rho^+
e^{-\xi}}{(\rho^++\rho^-)^2} \ ,
\nonumber \\
\tP^{\tau\tau}&=&4\frac{e^{-\xi}}{(\rho^++\rho^-)^2}
\ , \quad
\tP^{\sigma\sigma}=4\frac{\rho^+\rho^-e^{-\xi}}
{(\rho^++\rho^-)^2}  \ , \nonumber \\
\tP^{\tau\sigma}&=&4\frac{\rho^+e^{-\xi}}
{(\rho^++\rho^-)^2}  \ , \quad
\tP^{\sigma\tau}=-4\frac{\rho^-
e^{-\xi}}{(\rho^++\rho^-)^2} \ .
\nonumber \\
 \end{eqnarray}
Then we can easily find
the Hamiltonian density in the form
\begin{eqnarray}\label{mHflat}
\mH &=&
\partial_\tau x^mp_m+\partial_\tau \theta^\alpha
p_\alpha+\partial_\tau
\htheta^{\halpha} \hp_{\halpha}
-\mL= \nonumber \\
\nonumber \\
&=&-\frac{1}{\sqrt{-h}h^{\tau\tau}}[
\frac{1}{2}p_m\eta^{mn}p_n+\frac{1}{2}
\partial_\sigma
x^m\eta_{mn}\partial_\sigma x^n]
-\frac{h^{\tau\sigma}}{h^{\tau\tau}}
p_m\partial_\sigma x^m- \nonumber \\
&-&\frac{h_{\tau\tau}}{h_{\tau\sigma}+
\sqrt{-h}}p_\alpha \partial_\sigma
\theta^\alpha
-\frac{h_{\tau\tau}}{h_{\tau\sigma}-
\sqrt{-h}}\hp_{\halpha}
\partial_\sigma
\htheta^{\halpha}+
\nonumber \\
&+&\frac{h_{\tau\tau}}{h_{\tau\sigma}+
\sqrt{-h}}\pi_\alpha\partial_\sigma\lambda^\alpha+
\frac{h_{\tau\tau}}{h_{\tau\sigma}-
\sqrt{-h}}\hpi_{\halpha}\partial_\sigma\hlambda^{\halpha}
=\nonumber \\
&=&\frac{\rho^++\rho^-}{2}
\left[\frac{1}{2}p_m\eta^{mn}p_n+\frac{1}{2}
\partial_\sigma
x^m\eta_{mn}\partial_\sigma x^n\right]
+\frac{1}{2}(\rho^+-\rho^-)
p_m\partial_\sigma x^m
+\nonumber \\
&+& \rho_-(p_\alpha \partial_\sigma
\theta^\alpha-\pi_\alpha\partial_\sigma
\lambda^\alpha) -\rho_+ (\hp_{\halpha}
\partial_\sigma
\htheta^{\halpha}-\hpi_{\halpha}\partial_\sigma
\hlambda^{\halpha})=
 \nonumber \\
 &=&\rho^+T_++\rho^-T_- \ ,  \nonumber \\
\end{eqnarray}
where
\begin{eqnarray}\label{T+-}
T_-&=&\frac{1}{4}[ p_m\eta^{mn}p_n+
\partial_\sigma
x^m\eta_{mn}\partial_\sigma x^n]
-\frac{1}{2}p_m\partial_\sigma x^m
+p_\alpha \partial_\sigma
\theta^\alpha- \pi_\alpha
\partial_\sigma \lambda^\alpha \ ,
\nonumber \\
T_+&=&\frac{1}{4} [p_m\eta^{mn}p_n+
\partial_\sigma
x^m\eta_{mn}\partial_\sigma x^n]
+\frac{1}{2} p_m\partial_\sigma x^m
-\hp_{\halpha}
\partial_\sigma
\htheta^{\halpha}+\hpi_{\halpha}\partial_\sigma
\hlambda^{\halpha} \   \nonumber \\
\end{eqnarray}
or alternatively
\begin{eqnarray}
T_-&=&\frac{1}{4}
\eta^{mn}\Pi_{-m}\Pi_{-n}+d_\alpha\partial_\sigma
\theta^\alpha-\pi_\alpha\partial_\sigma\lambda^\alpha
 \ , \nonumber \\
T_+&=&\frac{1}{4}\eta^{mn} \Pi_{+m}
\Pi_{+n}
 -\hd_{\halpha}\partial_\sigma
\htheta^{\halpha} +\hpi_{\halpha}
\partial_\sigma \hlambda^{\halpha} \ ,
\nonumber \\
\end{eqnarray}
where
\begin{eqnarray}
\Pi_{+m}&=&(p_m+\partial_\sigma
x^n\eta_{nm} -2i (\htheta
\gamma^n\partial_\sigma
\htheta)\eta_{nm}) \ , \nonumber \\
\Pi_{-m}&=&(p_m-\partial_\sigma
x^n\eta_{nm}
+2i(\theta\gamma^n\partial_\sigma\theta)\eta_{nm})
\ .
\nonumber \\
\end{eqnarray}
Note that we  also
 used following relations
\begin{eqnarray}
d_{\alpha\mu}\cP^{\mu\sigma}&=&
\rho^- d_{\alpha\mu}\cP^{\mu\tau} \ ,
\nonumber \\
\hd_{\halpha\mu}\tP^{\mu\sigma}&=&-
\rho^+ d_{\halpha\mu}\tP^{\mu\tau} \
 \  \nonumber \\
\end{eqnarray}
that follow from the explicit form of
projectors given in (\ref{projectors})
\footnote{It is also clear that this
relations hold for further chiral
variables.}.
Now with the help of the standard
Poisson brackets that are collected
 in Appendix
we easily obtain
\begin{eqnarray}
\pb{d_{\alpha}(\sigma), T_-(\sigma')}
 &=&
-d_\alpha(\sigma)\partial_\sigma \delta(\sigma-\sigma')
-\partial_\sigma d_\alpha(\sigma)
\delta(\sigma-\sigma')
 \ ,
 \nonumber \\
\pb{\hd_{\halpha}(\sigma),T_+(\sigma')}
&=&\hd_{\halpha}(\sigma)
\partial_\sigma \delta(\sigma-\sigma')
+\partial_\sigma \hd_{\halpha}(\sigma)
 \delta(\sigma-\sigma') \ .
\nonumber \\
\end{eqnarray}
Then it is easy to  determine the
Poisson brackets between $Q$ and
$T_+,T_-$ and we obtain
\begin{eqnarray}\label{QT+-}
\pb{Q_L,T_+(\sigma)}=0 \ , \quad
\pb{Q_R,T_+(\sigma)}=0 \ , \nonumber \\
\pb{Q_L,T_-(\sigma)}=0 \ , \quad
\pb{Q_R,T_-(\sigma)}=0 \ . \nonumber \\
\end{eqnarray}
Further, after some algebra we
determine the Poisson bracket between
Virasoro constraints
\begin{eqnarray}
\pb{T_+(\sigma),T_+(\sigma')}&=&
2T_+(\sigma)\partial_\sigma
\delta(\sigma-\sigma')+
\partial_\sigma T_+(\sigma)
\delta(\sigma-\sigma') \ , \nonumber \\
\pb{T_-(\sigma),T_-(\sigma')}&=&
-2T_-(\sigma)\partial_\sigma
\delta(\sigma-\sigma')-
\partial_\sigma T_-(\sigma)
\delta(\sigma-\sigma') \ ,  \nonumber \\
\pb{T_+(\sigma),T_-(\sigma')}&=&0 \ .
\nonumber \\
\end{eqnarray}
We see that these Poisson brackets take
standard form. This fact will be
important bellow.

 Now we come to an important point in
pure spinor formalism. Although there
is no fundamental $b$ ghost in the pure
spinor formalism  one can construct a
composite operators $G_\pm$ that obey
the relations
\begin{equation}\label{QG}
\pb{Q,G_+(\sigma)}=T_+(\sigma) \ ,
\quad \pb{Q,G_+(\sigma)}=T_+(\sigma) \ ,
\end{equation}
where
\begin{eqnarray}\label{defGf}
G_+&=&\frac{\hC_{\halpha}
G^{\halpha}_+} {\hC_{\halpha}
\hlambda^{\halpha}} \ , \quad
G_-=\frac{C_\alpha G^\alpha_-} {C_\alpha
\lambda^\alpha} \ ,
\nonumber \\
G^{\halpha}_+&=&
-\frac{i}{8}\Pi^m_+(\gamma_m
\hd)^{\halpha}
+\frac{1}{4}\hN_{mn}(\gamma^{mn}\partial_\sigma
\htheta)^{\halpha}+\frac{1}{4}
\hJ\partial_\sigma \htheta^{\halpha}
\ , \nonumber \\
G^{\alpha}_-&=&
-\frac{i}{8}\Pi^m_-(\gamma_m
d)^{\alpha}
-\frac{1}{4}N_{mn}(\gamma^{mn}\partial_\sigma
\theta)^{\alpha}-\frac{1}{4}
J\partial_\sigma \theta^{\alpha}
\ . \nonumber \\
\end{eqnarray}
Alternatively, using the fact that
$\pb{Q,\lambda^\alpha}=0$  we can write
\begin{equation}\label{pbQG}
\pb{Q,G_-}=\frac{C_\alpha}{C_\alpha
\lambda^\alpha}\pb{Q,G^\alpha_-}=T_- \
.
\end{equation}
In fact, using Poisson brackets given
in Appendix and using   important
identities
\begin{eqnarray}
& &\delta_\beta^\gamma
\delta_\alpha^\delta =
\frac{1}{2}\gamma^m_{\alpha\beta}
\gamma_m^{\gamma\delta} -\frac{1}{8}
(\gamma^{mn})_\alpha^\gamma
(\gamma_{mn})^\delta_\beta-\frac{1}{4}
\delta_\alpha^\gamma
\delta_\beta^\delta \ , \nonumber \\
& &(\gamma^m)_{\gamma\delta}
(\gamma_m)_{\alpha\beta}+
(\gamma^m)_{\gamma\alpha}(\gamma_m)_{\beta\delta}+
(\gamma^m)_{\gamma\beta}(\gamma_m)_{\delta\alpha}=0
\nonumber \\
\end{eqnarray}
we can show that the Poisson brackets
between $Q$ and $G^\alpha_+$ given in
(\ref{defGf}) is equal to
\begin{equation}
 \pb{Q,G^\alpha_-}=\lambda^\alpha T_- \
\end{equation}
that confirms (\ref{pbQG}).
In the same way we obtain
\begin{eqnarray}
\pb{Q,G^{\halpha}_+}=\hlambda^{\halpha}T_+
\ .
\end{eqnarray}
As the next step we calculate the
Poisson bracket
$\pb{G_-(\sigma),G_-(\sigma')}$. The
calculation of this bracket is
 non-trivial and deserves
careful calculation. However after some
work we derive important result
\begin{equation}\label{G+G+}
\pb{G_-(\sigma),G_-(\sigma')}=0 \ .
\end{equation}
It is important to stress that the
Poisson bracket given above vanish on
condition that $C^\alpha$
is pure spinor: $C^\alpha
(\gamma^m)_{\alpha\beta}C^\beta=0$.
 In the same way we obtain that
\begin{equation}\label{G_G_}
\pb{G_+(\sigma),G_+(\sigma')}=0 \ .
\end{equation}
As we will see below the fact that the
Poisson brackets (\ref{G+G+}) and
(\ref{G_G_}) are zero has an important
consequence for the correct
redefinition of the BRST operator.

Further we determine
 the Poisson brackets between $T_\pm$
and $G_\pm$. Using the formulas
collected in Appendix we easily obtain
\begin{eqnarray}\label{pbT+G}
\pb{T_+(\sigma),G_+(\sigma')}&=&
-2G_+(\sigma)\partial_\sigma
\delta(\sigma-\sigma')-
\partial_\sigma G_+(\sigma)\delta(\sigma-\sigma') \ ,
\nonumber \\
\pb{T_-(\sigma),G_-(\sigma')}&=&
2G_-(\sigma)\partial_\sigma
\delta(\sigma-\sigma')+
\partial_\sigma G_-(\sigma)\delta(\sigma-\sigma') \ ,
\nonumber \\
\pb{T_\pm(\sigma),G_\mp(\sigma')}&=&0 \ . \nonumber \\
\end{eqnarray}
After these preliminary calculations
we follow \cite{Berkovits:2007wz} and
add to the BRST operator (\ref{QLR})
cohomology trivial
 term  $\int d\sigma(\tilde{\gamma}^+b_++\tilde{\gamma}^-
b_-)$ so that the BRST operator takes
the form
\begin{eqnarray}\label{Qnt}
 Q= \int d\sigma
(\lambda^\alpha d_{\alpha}+
\hlambda^{\halpha}\hd_{\halpha}
+\tilde{\gamma}^+b_++\tilde{\gamma}^-
b_-) \ , \nonumber \\
\end{eqnarray}
where
$(\tilde{\beta}_\pm,\tilde{\gamma}^\pm)$
are bosonic and $(b_\pm ,c^\pm)$ are fermionic
fields that have following Poisson
bracket  structure
\begin{eqnarray}\label{pbbc}
\pb{c^\pm(\sigma),b_\pm(\sigma')}=
-\delta(\sigma-\sigma') \ , \quad
\pb{\tilde{\gamma}^\pm(\sigma),
\tilde{\beta}_\pm(\sigma')}=\delta(\sigma-\sigma')
\ .
\end{eqnarray}
Using these Poisson bracket and the
form of the BRST operator (\ref{Qnt})
we easily determine the transformation
properties of these fields under BRST
transformations
\begin{eqnarray}
\pb{Q,c^+(\sigma)}&=&-\tilde{\gamma}^+(\sigma)
\ , \quad
\pb{Q,c^-(\sigma)}=-\tilde{\gamma}^-(\sigma)
\ ,
\nonumber \\
\pb{Q,\tilde{\beta}_+(\sigma)}&=&
b_+(\sigma) \ , \quad
\pb{Q,\tilde{\beta}_-(\sigma)}=
b_-(\sigma) \ . \nonumber \\
\end{eqnarray}
We also suggest  that these fields
contribute to the Hamiltonian density
as
\begin{eqnarray}
\delta H&\equiv& \int
d\sigma (\rho^+(-\tbeta_+\partial_\sigma
\tgamma^+-\partial_\sigma
(\tbeta_+\tgamma^+)+b_+\partial_\sigma
c^+ )+ \nonumber
\\
&+&\rho^-(\tbeta_-\partial_\sigma
\tgamma^-+\partial_\sigma
(\tbeta_-\tgamma^-) -b_-\partial_\sigma
c^-))\nonumber \\
\end{eqnarray}
since then the  time evolution of
$\tbeta,\tgamma, b,c$ has an expected
form
\begin{eqnarray}\label{eqghost}
\partial_\tau \tbeta_\pm &=&\pb{\delta H,\tbeta_\pm}=
\pm \partial_\sigma
(\rho^\pm\tbeta_\pm) \ , \nonumber \\
\partial_\tau \tgamma^\pm&=&
\pb{\delta
H,\tgamma^\pm}=\pm\rho^\pm\partial_\sigma
\tgamma^\pm \ , \nonumber \\
\partial_\tau c^\pm&=&\pb{\delta H,c^\pm
}=\pm \rho^\pm\partial_\sigma c^\pm \ .
\nonumber \\
\partial_\tau b_\pm &=&
\pb{\delta H,b^\pm}=
\pm \partial_\sigma (\rho^\pm b_\pm) \ ,
\nonumber \\
\end{eqnarray}
In what follows we return to the
standard presumption of the pure spinor
formalism that the world-sheet metric
is flat. Then $\rho^+=\rho^-=1$ and
using (\ref{pbT+G}) we easily determine
\begin{eqnarray}\label{eqG}
\partial_\tau G_\pm&=&
\pm \partial_\sigma G_\pm \ . \nonumber \\
\end{eqnarray}
Now we are ready to perform the
redefinition of the BRST operator
(\ref{Qnt}). Let us  consider an
operator
\begin{equation}
R=\int d\sigma
(c^+G_++c^-G_-+c^+\partial_\sigma
c^+\beta_{+}- c^-\partial_\sigma
c^-\beta_-) \ .
\end{equation}
Using (\ref{eqghost}) and (\ref{eqG})
it is easy to see that  $R$ is
conserved
\begin{eqnarray}
\partial_\tau R=
\int d\sigma \partial_\sigma (\dots)=0
\ ,
\nonumber \\
\end{eqnarray}
where we implicitly presume that the
world-sheet modes obey appropriate
boundary conditions.
%
Our goal is to perform classical
analogue of the  redefinition of the
BRST operator $Q$ that was performed in
\cite{Berkovits:2007wz}. In order to
clearly understand of this redefinition
in the context of classical mechanics
we will be
 slightly formal and
consider either matrix valued functions
or  quantum mechanics operators $F,Q$ and
$R$  in the
form
\begin{equation}
 F(x) = e^{xR}Q(e^{-xR}) \ ,
 \end{equation}
 where $x$ is free parameter.

 As the next step we make
  an expansion around the
point $x=0$ so that
\begin{equation}
 F(x)=\sum_{n=0}^\infty
 \frac{1}{n!}\frac{d^nF}{d^nx}(0)x^n
 \end{equation}
 and where
\begin{eqnarray}
\frac{dF}{dx}(0)&=&
[R,Q]
\ , \nonumber \\
\frac{d^2F}{d^2x}(0)&=& [R,[R,Q]] \ ,
\nonumber \\
\dots \nonumber \\
\frac{d^nF}{d^nx}(0)&=& \overbrace{
[R,\dots,[R,[R,Q]]]}^n \ .
\end{eqnarray}
Then putting $x=1$ and using the fact
that $F(x=1)=Q'$
we obtain the formal
expression for $Q'$ in the form
\begin{equation}
Q'=
 Q+\sum_{n=1}^n \frac{1}{n!}
\overbrace{ [R,\dots,[R,[R,Q]]]}^n \ .
\end{equation}
Using the analogy between Poisson
brackets in classical mechanics and
commutators in quantum mechanics we
propose the classical redefinition of
the operator $Q'$ in the form
\begin{equation}
Q'= Q+\sum_{n=1}^n \frac{1}{n!}
\overbrace{ \pb{R,\dots,\pb{R,\pb{R,Q}}}}^n
\end{equation}
where  $\pb{\dots}$ corresponds to
graded Poisson bracket. In usual situation
the sum above terminates after few
steps. Before we proceed to the
explicit determination of $Q'$ we show
that the new operator $Q'$ is
conserved:$\frac{d}{d\tau}Q'=0$. In
fact, since $\partial_\tau R=0$ we
easily obtain that
\begin{equation}
\frac{dQ'}{d\tau}=0 \ .
\end{equation}

Now we proceed to the explicit
calculation of $Q'$. To do this we have
to calculate the Poisson brackets
$\pb{R,Q},\pb{R,\pb{R,Q}},\dots$.
Firstly,  we have
\begin{eqnarray}
\pb{R,Q}&=&
\int d\sigma (c^+(
T_+-\tbeta^+\partial_\sigma \tgamma^+
-\partial_\sigma (\tbeta_+\tgamma^+)
+b_+\partial_\sigma
c^+)+\nonumber \\
&+& c^-(T_-+\tbeta_-\partial_\sigma
\tgamma^-+
\partial_\sigma (\tbeta_-\tgamma^-)
-b_-\partial_\sigma c^-)
+\tgamma^+G_++\tgamma^-G_-) \ ,
\nonumber \\
\end{eqnarray}
where we used
\begin{eqnarray}\label{pbRc}
\pb{R,c^\pm}&=&0 \ , \quad
\pb{R,\tbeta^\pm}=0
 \nonumber
\\
\pb{R,\tgamma^\pm}&=&\mp c^\pm
\partial_\sigma c^\pm \ ,
\nonumber \\
\pb{R,b^+}&=&G^++\partial_\sigma
c^+ \tbeta_++\partial_\sigma
(c^+\tbeta_+) \ , \nonumber \\
\pb{R,b^-}&=&G^--
\partial_\sigma c^-\tbeta^-
-\partial_\sigma (c^-\tbeta_-)
\nonumber \\
\end{eqnarray}
As the next step we calculate
$\pb{R,\pb{Q,R}}$. In fact, using
(\ref{pbRc})  we  obtain the result
\begin{equation}
\pb{R,\pb{R,Q}}=0 \ .
\end{equation}
In other words we obtain the BRST operator $Q'$ in
the form
\begin{eqnarray}
Q'&=&\int d\sigma
[c^+\tilde{T}_++c^-\tilde{T}_-+
\lambda^\alpha d_{\alpha}+
\hlambda^{\halpha}\hd_{\halpha}+
\nonumber \\
&+&\tgamma^+G_++\tgamma^-G_-+
\tgamma^+b_++\tgamma^-b_-] \ ,  \nonumber \\
\end{eqnarray}
where
\begin{eqnarray}
\tilde{T}_+&=&
T_+-\tbeta^+\partial_\sigma \tgamma^+
-\partial_\sigma (\tgamma^+\tbeta_+)
+b_+\partial_\sigma
c^+ \ , \nonumber \\
\tilde{T}_-&=&
T_-+\tbeta_-\partial_\sigma \tgamma^-+
\partial_\sigma (\tbeta_-\tgamma^-)
-b_-\partial_\sigma c^- \ .  \nonumber \\
\end{eqnarray}
 This is the standard form of the BRST
operator for closed superstring when we
interpret
$(\tgamma^+,\lambda^+,\lambda_{ab},\tgamma^-,
\hlambda^+,\hlambda_{ab})$ as $24$
independent bosonic ghosts together
with two sets of Virasoro constraints.
It was shown in \cite{Berkovits:2007wz}
 that this action is closely related
to Green-Schwarz superstring. It is also
nice to see that the new BRST operator
contains Virasoro constraints whose presence
was hidden in the original formulation
of pure spinor string.
\section{General Background}\label{third}
In this section we extend the
discussion presented in previous
section to the case of pure spinor
string in general background
\footnote{We omit the Fradkin-Tseytlin
term $\int \Phi(Z)r$ where $\Phi$ is
dilaton super-field and $r$ is
world-sheet curvature.}. Recall that
this action takes the form
\begin{eqnarray}\label{genactcov}
S&=&-\int d^2\sigma\sqrt{-h} (\frac{1}{2}
h^{\mu\nu}g_{\mu\nu}
-\frac{1}{2}\epsilon^{\mu\nu} b_{\mu\nu}
+ \nonumber \\
&+& P^{\alpha\hbeta}d_{\alpha\mu}
 \cP^{\mu\nu}
\hd_{\hbeta\nu}+d_{\alpha\mu}
\cP^{\mu\nu} \partial_\nu
Z^ME^\alpha_M(Z)
 +\hd_{\halpha\mu}
\tP^{\mu\nu}\partial_\nu Z^M
E^{\halpha}_M(Z) + \nonumber \\
&+& w_{\mu\beta}\lambda^\alpha
\cP^{\mu\nu}\partial_\nu
Z^M\Omega_{M\alpha}^{ \ \beta}+
 \hw_{\mu\hbeta}\hlambda^{\halpha}
\tP^{\mu\nu}
\partial_\nu Z^M
\hat{\Omega}_{M\halpha}^{\hbeta} +
 C_{\alpha}^{\beta \hat{\gamma}}
\lambda^\alpha w_{\beta\mu}
 \tP^{\mu\nu}
\hd_{\hgamma\nu}+
\nonumber \\
&+& \hat{C}^{\hbeta\gamma}_{\halpha}
\hlambda^{\halpha}\hw_{\hbeta\mu}
\cP^{\mu\nu}d_{\gamma\nu}+ S^{\beta
\hdelta}_{\alpha \hgamma}
\lambda^\alpha w_{\beta\mu} \cP^{\mu\nu}
 \hlambda^{\halpha}\hw_{\hdelta\nu})
 +S^\lambda+ S^{\hlambda} \ ,
\nonumber \\
\end{eqnarray}
where
\begin{eqnarray}
S^\lambda &=&-\int d^2\sigma
\sqrt{-h}
w_{\mu\alpha}\cP^{\mu\nu}
\partial_\nu\lambda^\alpha
 \ , \quad  S^{\hlambda}= -\int
d^2\sigma\sqrt{-h}
\hw_{\mu\halpha}\tP^{\mu\nu}\partial_\nu
\hlambda^{\halpha}\ . \nonumber \\
\end{eqnarray}
Note also that $g_{\mu\nu}$ and
$b_{\mu\nu}$ that appear in
(\ref{genactcov}) are defined as
\begin{equation}
G_{\mu\nu}=\partial_\mu Z^ME_M^a
\partial_\nu Z^NE_N^b\eta_{ab} \ ,
\quad
b_{\mu\nu}=
\partial_\mu Z^M\partial_\nu Z^Nb_{MN} \ ,
\end{equation}
and where $M=(m,\mu,\hmu)$ are
curved superspace indices,
$Z^M=(x^m,\theta^\mu \ ,
\theta^{\hmu})$, $A=(a,\alpha,\halpha)$
are tangent superspace indices,
$S^\lambda \ , \ S^{\hlambda}$ are the
flat actions for the pure spinor
variables. Finally
$E_M^\alpha \ , \quad E_M^{\halpha},
\quad  \Omega_{M\alpha}^{ \ \beta} \ ,
\quad \hat{\Omega}_{M\halpha}^{\
\hbeta} \ , \quad  P^{\alpha\hbeta} \ ,
\quad C_{\alpha}^{\beta\hgamma} \ ,
\quad \hat{C}^{\hbeta\gamma}_{\halpha}
\ , \quad S^{\beta
\hdelta}_{\alpha\hgamma} $
are background space-time fields. Note
also  that $d_{\mu\alpha} \ ,
\hd_{\mu\halpha}$ should be treated as
 independent variables since
$p_\alpha \ , \hat{p}_{\halpha}$ do not
appear explicitly in the action.

As in the flat space the fundamental
object of  the pure spinor formalism in
the general background  is
the BRST operator $Q=Q_L+Q_R$
where
\begin{eqnarray}\label{QLRg}
Q_L&=&\int d\sigma \lambda^\alpha
d_{\alpha\mu}\sqrt{-h}\cP^{\mu \tau} \ ,
\quad
Q_R=\int d\sigma\hlambda^{\halpha}
\hd_{\halpha\mu}\sqrt{-h}
\tP^{\mu \tau} \ .
\nonumber \\
\end{eqnarray}
Properties of these operators
were carefully studied in
\cite{Berkovits:2001ue} and we
recommend this paper for more details.

In order to use the classical formalism
we have to  express $d_{\mu\alpha} \ ,
\hd_{\mu\halpha}$ in terms of  the
canonical variables of the extended
phase space spanned by coordinates
$ (Z^M,\lambda^\alpha,
\hlambda^{\halpha},
P_M,\pi_{\alpha},\hpi_{\halpha})$ where
\begin{eqnarray}\label{PMg}
P_M&=&\frac{\delta S}{\delta
\partial_\tau Z^M}=
-\sqrt{-h}h^{\tau\mu}
E_M^a \eta_{ab}\partial_\mu Z^N
E^b_N
+\partial_\sigma Z^N b_{MN}+\nonumber \\
&+& \sqrt{-h}[E_M^\alpha d_{\alpha\mu} \cP^{\mu \tau}+
 E^{\halpha}_M d_{\halpha \mu}
\tP^{\mu \tau}-
\nonumber \\
&-&w_{\mu\beta}\lambda^\alpha \cP^{\mu
\tau}\Omega_{M\alpha}^{ \
\beta}-\hw_{\mu\hbeta}\hlambda^{\halpha}
\tP^{\mu \tau}
\hat{\Omega}_{M\halpha}^{\hbeta}] \
 \nonumber \\
\end{eqnarray}
and
\begin{eqnarray}
\pi_{\alpha}=\frac{\delta S}{\delta
\partial_\tau \lambda^\alpha}=
-w_{\mu\alpha}\sqrt{-h}\cP^{\mu\tau}\ , \quad
 \hpi_{\halpha}=\frac{\delta S}{\delta
\partial_\tau \hlambda^{\halpha}}=
-\hw_{\mu\halpha}\sqrt{-h}\tP^{\mu
\tau} \ .
\nonumber \\
\end{eqnarray}
By definition these momenta obey the
canonical graded Poisson brackets
\begin{eqnarray}\label{cpb}
\pb{Z^M(\sigma),P_N(\sigma')}&=&
(-1)^{|M|}\delta^M_N\delta(\sigma-\sigma') \ , \nonumber \\
\pb{\lambda^\alpha(\sigma),\pi_\beta(\sigma')}&=&
\delta_\beta^\alpha
\delta(\sigma-\sigma') \ , \quad
\pb{\hlambda^{\halpha}(\sigma),\hpi_{\hbeta}(\sigma')}=
\delta_{\hbeta}^{\halpha}
\delta(\sigma-\sigma') \ .
\nonumber \\
\end{eqnarray}
As the next step we  express
$d_{\alpha\mu}$ as functions of
canonical variables. To begin with we
use the definition of vielbein
\begin{equation}
E_A^ME_M^B=\delta_A^B \ .
\end{equation}
Then
\begin{equation}
E_\alpha^ME_M^b=0  \ , \quad
E_\alpha^ME_M^{\halpha}=0 \
\end{equation}
and consequently when we multiply (\ref{PMg}) with
$E^\alpha_M$ from the left
we can express
$d_{\alpha\mu}\cP^{\mu\tau},
\hd_{\halpha\mu}\tP^{\mu\tau}$ as
functions of  canonical variables
\begin{eqnarray}
d_{\alpha \mu}\cP^{\mu \tau}&=&
\frac{1}{\sqrt{-h}} E_\alpha^M[
P_M-\partial_\sigma Z^N
b_{MN}-\Omega_{M\gamma}^{ \ \beta}
\lambda^{\gamma}\pi_\beta -
\hat{\Omega}_{M\halpha}^{\hbeta}
\hlambda^{\halpha}\hpi_{\hbeta}]
 \equiv \frac{1}{\sqrt{-h}}d_\alpha \ ,
 \nonumber \\
\hd_{\mu\hbeta}\tP^{\mu\tau}&=&
\frac{1}{\sqrt{-h}}E_{\halpha}^M[
P_M-\partial_\sigma Z^N
b_{MN}-\Omega_{M\gamma}^{ \ \beta}
\lambda^{\gamma}\pi_\beta -
\hat{\Omega}_{M\halpha}^{\hbeta}
\hlambda^{\halpha}\hpi_{\hbeta}]
 \equiv \frac{1}{\sqrt{-h}}
\hd_{\halpha} \ .
\nonumber \\
\end{eqnarray}
It is also useful to introduce the
notation
\begin{eqnarray}
\Pi^{a}_\mu&=&\partial_\mu Z^ME_M^{a} \
, \quad  P_{a}=E_{a}^MP_M \  \ ,
\nonumber \\
\Pi_{\tau}^a&=&
 -\frac{1}{\sqrt{-h}h^{\tau\tau}}\eta^{ab}
\hat{P}_b-\frac{h^{\tau\sigma}}{h^{\tau\tau}}
\Pi_\sigma^a \ ,  \nonumber \\
\end{eqnarray}
where
\begin{equation}
\hat{P}_A=
E_A^MP_M-E_A^M\partial_\sigma Z^N
b_{MN}-E_A^M \Omega_{M\beta}^\alpha
\pi_\alpha \lambda^\beta-E_A^M
\hOmega_{M\hbeta}^{\halpha}\hpi_{\halpha}
\hlambda^{\hbeta} \ .
\end{equation}
With this notation and after some
 work we derive the Hamiltonian
density for pure spinor string in general
background in the form
\begin{eqnarray}
\mH &=&\partial_\tau\lambda^\alpha\pi_\alpha+
\partial_\tau \hlambda^{\halpha}
\hpi_{\halpha}+\partial_\tau Z^MP_M-\mL =
\nonumber \\
&=& -
\frac{1}{2}\sqrt{-h}h^{\tau\tau}
\Pi^{a}_\tau
\Pi^{b}_\tau\eta_{ab}
+\frac{1}{2}\sqrt{-h}h^{\sigma\sigma}
\Pi_{\sigma}^{a} \Pi_\sigma^{b}
\eta_{ab}
+\nonumber\\
&+&\frac{1}{2}\sqrt{-h}[ C_{\alpha}^{\beta \hat{\gamma}}
\lambda^\alpha w_{\beta\mu}
 \tP^{\mu\rho_1}h_{\rho_1\rho_2}
 \tP^{\rho_2\nu}
\hd_{\hgamma\nu}+\nonumber \\
&+& \hat{C}^{\hbeta\gamma}_{\halpha}
\hlambda^{\halpha}\hw_{\hbeta\mu}
\cP^{\mu\rho_1}h_{\rho_1\rho_2}\cP^{\rho_2\nu}
d_{\gamma\nu}+ S^{\beta
\hdelta}_{\alpha \hgamma}
\lambda^\alpha w_{\beta\mu}
\cP^{\mu\rho_1}
h_{\rho_1\rho_2}\cP^{\rho_2\nu}
 \hlambda^{\halpha}\hw_{\hdelta\nu}]+
 \nonumber \\
 &+&\sqrt{-h}[w_{\mu\alpha}\lambda^\beta \cP^{\mu\sigma}
\partial_\sigma Z^M\Omega_{M\alpha}^\beta+
+\hw_{\mu\halpha}\hlambda^{\hbeta}
\tP^{\mu\sigma}\hOmega_{M\halpha}^{\hbeta}
+
 \nonumber \\
&+& d_{\alpha\mu}\cP^{\mu\sigma}
\Pi_\sigma+d_{\halpha\mu}\tP^{\mu\sigma}
\Pi_\sigma+ P^{\alpha\hbeta}
d_{\alpha\mu}\cP^{\mu\nu}d_{\hbeta\nu}]
\nonumber \\
&+&\sqrt{-h}w_{\mu\alpha} \cP^{\mu\sigma}
\partial_\sigma\lambda^\alpha+
\sqrt{-h}\hw_{\mu\halpha}
\tP^{\mu\sigma}\partial_\sigma\hlambda^{\hbeta}
\equiv  \rho^+T_++\rho^-T_- \ ,  \nonumber \\
\end{eqnarray}
where
\begin{eqnarray}
T_+&=&\frac{1}{4}
(\hP_a+\eta_{ac}\Pi^c_\sigma)\eta^{ab}
(\hP_b+\eta_{bd}\Pi^d_\sigma)-\hd_{\halpha}
\Pi^{\halpha}_\sigma
-\frac{1}{2}(\pi_\beta\lambda^\alpha C_\alpha^{\beta\hgamma}
\hd_{\hgamma}+\hpi_{\hbeta}\hlambda^{\halpha}
C_{\halpha}^{\hbeta\gamma}d_{\gamma})+
\nonumber \\
&+&\frac{1}{2}S_{\alpha\hgamma}^{\beta\hdelta}
\pi^\alpha \lambda_\beta
\hlambda^{\hgamma}\pi_{\hdelta}+\frac{1}{2}
P^{\alpha\hbeta}d_\alpha
d_{\hbeta}+\hpi_{\halpha}\partial_\sigma \hlambda^{\halpha} \ ,
\nonumber \\
T_-&=&\frac{1}{4}
(\hP_a-\eta_{ac}\Pi^c_\sigma)\eta^{ab}
(\hP_b-\eta_{bd}\Pi^d_\sigma)+d_{\alpha}
\Pi^{\alpha}_\sigma
+\frac{1}{2}(\pi_\beta\lambda^\alpha C_\alpha^{\beta\hgamma}
\hd_{\hgamma}+\hpi_{\hbeta}\hlambda^{\halpha}
C_{\halpha}^{\hbeta\gamma}d_{\gamma})-
\nonumber \\
&-&\frac{1}{2}S_{\alpha\hgamma}^{\beta\hdelta}
\pi^\alpha \lambda_\beta
\hlambda^{\hgamma}\hpi_{\hdelta}-
\frac{1}{2}P^{\alpha\hbeta}d_\alpha
d_{\hbeta}-\pi_\alpha\partial_\sigma
\lambda^\alpha \ .
\nonumber \\
\end{eqnarray}
Following the logic of previous section
it seams to be  natural to determine
Poisson brackets structure between
$T_\pm$'s. However this is very
difficult task and the resulting
Poisson brackets are not very
interesting \footnote{For careful
discussion of this problem in the
context of Green-Schwarz superstring in
general background see
\cite{Shapiro:1989hd}.}.
Then in
order to derive some useful results and
predictions we  presume that the
Poisson brackets of the Virasoro
components $T_\pm$ take standard form
\begin{eqnarray}\label{PBvir}
\pb{T_+(\sigma),T_+(\sigma')}&=&
-2T_+(\sigma)\partial_\sigma
\delta(\sigma-\sigma')-\partial_\sigma
T_+(\sigma)\delta(\sigma-\sigma') \ ,
\nonumber \\
\pb{T_-(\sigma),T_-(\sigma')}&=&
2T_-(\sigma)\partial_\sigma
\delta(\sigma-\sigma')+\partial_\sigma
T_-(\sigma)\delta(\sigma-\sigma') \ ,
\nonumber \\
\pb{T_+(\sigma),T_-(\sigma')}&=& 0 \ .
\nonumber \\
\end{eqnarray}
Let us again introduce two
objects $G_\pm$ defined as
\begin{eqnarray}
\pb{Q,G_\pm(\sigma)}=T_\pm(\sigma) \ .
\end{eqnarray}
Then  using the nilpotence of $Q$:$\pb{Q,Q}=0$
 we obtain
\begin{equation}
\pb{Q,T_+(\sigma)}=0 \ , \quad
\pb{Q,T_-(\sigma)}=0 \ .
\end{equation}
As the next step we analyze the Poisson
brackets between $T_\pm$ and $G_\pm$.
Without knowledge of explicit form of
$G_\pm$ we guess their forms as
\begin{eqnarray}\label{TGpro+}
\pb{T_+(\sigma),G_+(\sigma')}&=&
-2G_+(\sigma)\partial_\sigma \delta(\sigma-\sigma')-
\partial_\sigma G_+(\sigma)\delta(\sigma-\sigma') \ ,
  \nonumber \\
\pb{T_-(\sigma),G_+(\sigma')}&=&
2G_-(\sigma)\partial_\sigma \delta(\sigma-\sigma')+
\partial_\sigma G_-(\sigma)\delta(\sigma-\sigma') \
. \nonumber \\
\end{eqnarray}
To check that this is correct proposal
we calculate the Poisson bracket
of $Q$ with left and right side
of the first equation in (\ref{TGpro+}). The Poisson
bracket of left-side with $Q$
gives
\begin{eqnarray}
& &\pb{Q,\pb{T_+(\sigma),G_+(\sigma')}}=
\pb{T_+(\sigma),\pb{Q,G_+(\sigma')}}
-\pb{G_+(\sigma'),\pb{Q,T_+(\sigma)}}=
\nonumber \\
&=&\pb{T_+(\sigma),\pb{Q,G_+(\sigma')}}=
\pb{T_+(\sigma),T_+(\sigma')}
\nonumber \\
\end{eqnarray}
while the Poisson bracket of $Q$
with right-side of  (\ref{TGpro+})
gives
\begin{eqnarray}
& &-2\pb{Q,G_+(\sigma)}\partial_\sigma
\delta(\sigma-\sigma')-
\partial_\sigma \pb{Q,G_+(\sigma)}\delta(\sigma-\sigma')=
\nonumber \\
&=&-2T_+(\sigma)\partial_\sigma
\delta(\sigma-\sigma')-
T_+(\sigma)\delta(\sigma-\sigma') \ .
\nonumber \\
\end{eqnarray}
Collecting these results we derive the
Poisson brackets (\ref{PBvir}). In the same
way we can proceed with the second equation
in (\ref{PBvir}).

Further, let us consider the Poisson
bracket $\pb{G_+(\sigma),T_-(\sigma')}$
and apply BRST operator $Q$ on it
\begin{eqnarray}
\pb{Q,\pb{G_+(\sigma),T_-(\sigma')}}=
\pb{T_-(\sigma'),\pb{Q,G_+(\sigma)}}-\pb{G_+(\sigma),
\pb{Q,T_-(\sigma')}}=0 \nonumber \\
\end{eqnarray}
using the fact that
$\pb{T_+(\sigma),T_-(\sigma')}=
\pb{Q,T_-(\sigma')}=0$. Consequently we
generally have
\begin{eqnarray}\label{TG-+}
\pb{T_-(\sigma),G_+(\sigma')}&=&
\pb{Q,\Omega_{-+}(\sigma,\sigma')} \ ,
\nonumber \\
\pb{T_+(\sigma),G_-(\sigma')}&=&
\pb{Q,\Omega_{-+}(\sigma,\sigma')} \
\end{eqnarray}
for some ghost number $-2$ functions
$\Omega_{-+}$ and $\Omega_{+-}$.
However the fact that the Poisson
bracket between $T_\pm$ and $G_\mp$ is
non-zero has impact on  time evolution
of $G_+$ since
\begin{eqnarray}\label{ptG}
\partial_\tau G_+&=&
\pb{H,G_+}=
\partial_\sigma (\rho^+G_+)+ \pb{Q,
\mathbf{\Omega}_{+-}}
 \ , \nonumber \\
\partial_\tau G_-&=&
\pb{H,G_-}
=-\partial_\sigma (\rho^-G_-)
+\pb{Q,\mathbf{\Omega}_{-+}} \
,
\nonumber \\
\end{eqnarray}
 where
\begin{equation}
\mathbf{\Omega}_{-+}(\sigma)=
\int d\sigma'
\Omega_{-+}(\sigma,\sigma') \ , \quad
\mathbf{\Omega}_{+-}(\sigma)= \int
d\sigma'
\Omega_{+-}(\sigma,\sigma') \ .
\end{equation}
 This result
has an important consequence for time
evolution of the operator $R$ defined as
\begin{equation}
R=\int d\sigma
(c^+G_++c^-G_-+c^+\partial_\sigma
c^+\beta_{+}- c^-\partial_\sigma
c^-\beta_-) \
\end{equation}
since  using (\ref{ptG}) we easily
determine that $R$ is conserved up the
BRST invariant term
\begin{eqnarray}\label{ptauR}
\partial_\tau R&=&
\pb{Q,\mathbf{R}} \ , \nonumber \\
\mathbf{R}&=&\int d\sigma
[c^+(\sigma)\mathbf{\Omega}_{+-}(\sigma)+c^-(\sigma)
\mathbf{\Omega}_{-+}(\sigma)] \ .
\nonumber \\
\end{eqnarray}
Now we consider the Poisson brackets
between $G_{\pm}(\sigma)$'s. We
generally presume that they are
non-zero and take the form
\begin{eqnarray}\label{pbGAB}
\pb{G_A(\sigma),G_B(\sigma')}=
\bG_{AB}(\sigma,\sigma') \ , \quad
\bG_{AB}(\sigma,\sigma')=\bG_{BA}(\sigma',\sigma)
\ ,  \nonumber \\
\end{eqnarray}
where we used the notation $G_A \
,A=\pm$. For reasons outlined
above  it is hard to
determine the concrete form of the
matrix $\bG_{AB}$ from the first
principles. However let us apply the
BRST operator $Q$ on (\ref{pbGAB}) for
$A=+,B=+$. Then, with the help of
generalized Jacobi identity and using
(\ref{TGpro+}) we obtain
\begin{eqnarray}
\pb{Q,\pb{G_+(\sigma),G_+(\sigma')}}=0
\nonumber \\
\end{eqnarray}
and hence
\begin{equation}
\pb{Q,\bG_{++}(\sigma,\sigma')}=0
\end{equation}
that implies
\begin{equation}
\bG_{++}(\sigma,\sigma')
=\pb{Q,\bH_{++}(\sigma,\sigma')} \
\end{equation}
for some function
$\bH_{++}(\sigma,\sigma')$ of the ghost
number $-3$.
In the same way we obtain
\begin{equation}
\pb{G_-(\sigma),G_-(\sigma')}=
\bG_{--}(\sigma,\sigma') \ , \quad
\bG_{--}=\pb{Q,\bH_{--}(\sigma,\sigma')}
\ .
\end{equation}
Finally we apply $Q$ on
$\pb{G_+(\sigma),G_-(\sigma')}$ and we
obtain
\begin{eqnarray}\label{G+-}
&
&\pb{Q,\pb{G_+(\sigma),G_-(\sigma')}}=
\pb{T_-(\sigma'),G_+(\sigma)}+
\pb{T_+(\sigma),G_-(\sigma')}=
\nonumber \\
& & =\Omega_{-+}(\sigma',\sigma)+
\Omega_{+-}(\sigma,\sigma') =
\pb{Q,\bG_{+-}(\sigma,\sigma')} \ .
\nonumber \\
\end{eqnarray}
We see that  generally it is not possible
to write
$\bG_{+-}$ as $\pb{Q,\bH_{+-}}$. We
return to this issue below.

Now we are ready to discuss the
redefinition of $Q$ as in the previous
section. Following the same logic as there we
write $Q'$ as
\begin{equation}
Q'= Q+\sum_{n=1}^n \frac{1}{n!}
\overbrace{
\pb{R,\dots,\pb{R,\pb{R,Q}}}}^n \ .
\end{equation}
Now we will argue that-as opposite to
the case studied in previous
section-there are some subtleties with this
redefinition when the string is moving
in general background. In fact, due to
the result (\ref{ptauR}) it is not
completely clear that $Q'$ is time
independent as well.
To see this we  again consider a
quantum mechanics example and calculate
\begin{eqnarray}
\frac{dQ'}{d\tau}&=&\frac{d
e^R}{d\tau}Qe^{-R}+e^R\frac{dQ}{d\tau}e^{-R}
+e^R Q \frac{de^{-R}}{d\tau}
=\nonumber \\
&=&\frac{d e^R}{d\tau}Qe^{-R} -e^RQ
e^{-R}\frac{d e^R}{d\tau} e^{-R} \ ,
\nonumber \\
\end{eqnarray}
where we used $\frac{dQ}{d\tau}=0 \ ,
\frac{de^{-R}}{d\tau}=-e^{-R}\frac{de^R}{d\tau}
e^{-R}$. We see from the expression
above that in order  $Q'$ to be
time-independent we have to demand
$[R,\partial_\tau R]=0$. Then
$\frac{de^R}{d\tau}= e^R
\partial_\tau R$ and hence
\begin{eqnarray}
\frac{dQ'}{d\tau}&= & e^R \partial_\tau
R Q e^{-R}-e^R Q\partial_\tau R e^{-R}=
\nonumber \\
&=&e^R([Q,\mathbf{R}]Q-Q[Q,\mathbf{R}])e^{-R}=
e^R[[Q,\mathbf{R}],Q]e^{-R}=0 \ ,
\nonumber \\
\end{eqnarray}
where in the final step we used
$\frac{dR}{d\tau}=[Q,\mathbf{R}]$.
With the help of the example given
above we  now return to the classical
mechanics. We again presume that
\begin{equation}\label{RtR}
\pb{R,\partial_\tau R}=
\pb{R,\pb{Q,\mathbf{R}}}=0 \ .
\end{equation}
Then we get
\begin{equation}
\frac{dQ'}{d\tau}=
e^R\left(\pb{\partial_\tau R,Q}\right)e^{-R}=
e^R\left(\pb{\pb{Q,\mathbf{R}},Q}\right)e^{-R}=0
\end{equation}
and hence we obtain that $Q'$ is
conserved as well.
%
%
Let us now calculate explicit form of
$Q'$. It is easy to see that
\begin{eqnarray}
\pb{R,Q}&=& \int d\sigma (c^+(
T_+-\tbeta^+\partial_\sigma \tgamma^+
-\partial_\sigma (\tgamma^+\tbeta_+)
+b_+\partial_\sigma
c^+)+\nonumber \\
&+& c^-(T_-+\tbeta_-\partial_\sigma
\tgamma^-+
\partial_\sigma (\tgamma^-\tbeta_-)
-b_-\partial_\sigma c^-)
+\tgamma^+G_++\tgamma^-G_-) \ .
\nonumber \\
\end{eqnarray}
We see that this form coincides with
the form of the BRST operator $Q'$
derived in previous section. Let us
then calculate the second Poisson
bracket $\pb{R,\pb{R,Q}}$. It is clear
that the difference with respect to the
calculation presented in previous
section comes from the possible
non-trivial form of the Poisson
brackets $\pb{T_\pm,G_\mp}$ and
$\pb{G_A,G_B}$. Then after some
calculations we obtain
\begin{eqnarray}
& &\pb{R,\pb{R,Q}}= \pb{Q,\Sigma}+
\nonumber \\
&+&\int d\sigma d\sigma'
[(c^+(\sigma)\tgamma^-(\sigma')+
c^-(\sigma)\tgamma^+(\sigma'))
\bG_{-+}(\sigma,\sigma')] \ , \nonumber
\\
\end{eqnarray}
where
\begin{eqnarray}
\Sigma=\int d\sigma d\sigma'
c^-(\sigma)c^+(\sigma')
[\Omega_{+-}(\sigma',\sigma)-
\Omega_{-+}(\sigma,\sigma')+\nonumber
\\
+c^+(\sigma)\tgamma^+(\sigma')
\bH_{++}(\sigma,\sigma')+
c^-(\sigma)\tgamma^-(\sigma')
\bH_{--}(\sigma,\sigma')] \ .  \nonumber \\
\end{eqnarray}
On the other hand, as we argued above,
in order to find time-independent BRST
operator $Q'$ the operator $R$ should
obey the relation (\ref{RtR}). Let us
presume that $\partial_\tau R=0$ so
that
\begin{equation}
\Omega_{-+}=\Omega_{+-}=0 \ .
\end{equation}
Then (\ref{G+-}) implies
\begin{equation}
\bG_{+-}(\sigma,\sigma')=
\pb{Q,\bH_{+-}(\sigma,\sigma')}
\end{equation}
and consequently
\begin{equation}
 \pb{R,\pb {R,Q}}= \pb{Q,\Sigma'} \ ,
\end{equation}
where now
\begin{eqnarray}
\Sigma'&=& \int d\sigma d\sigma' [
c^+(\sigma)\tgamma^+(\sigma')
\bH_{++}(\sigma,\sigma')+
c^-(\sigma)\tgamma^-(\sigma')
\bH_{--}(\sigma,\sigma')+\nonumber \\
&+&(c^+(\sigma)\tgamma^-(\sigma')+
c^-(\sigma)\tgamma^+(\sigma'))\bH_{-+}
(\sigma,\sigma')] \ .  \nonumber \\
\end{eqnarray}
These results imply that the new BRST
operator $Q'$ contains additional terms
as opposite to the BRST operator in
flat space-time. This is a natural
consequence of the form of the Poisson
brackets (\ref{TG-+}) and
(\ref{pbGAB}). In fact, if the new BRST
operator $Q'$ is interpreted as the
standard BRST operator that contains
the first-class constraints only the
fact that the Poisson brackets
(\ref{TG-+}),(\ref{pbGAB}) are
non-trivial implies that there are
additional constraints that should be
taken into account.
 Moreover, the new
form of the BRST operator is not the
convention-looking one that is
constructed from the first-class
constraints only and with corresponding
structure constants. In fact, in order
to find such a form of the BRST
operator we have to presume that all
functions $\bG_{AB}$  vanish. In this
case the new BRST operator takes the
same form as the BRST operator in flat
space time with difference that $T_\pm$
and $G_\pm$ are defined for pure spinor
string in general background. In other
words on condition given above we
derive conventional-looking BRST
operator in general background that is
constructed from the first-class
constraints only. This result then
opens an interesting possibility to
study the classical solution of the pure
spinor string in general background
since the new BRST operator contains
Virasoro constraints that are crucial
for correct physical interpretations of
these solutions.
\section{Conclusion}\label{fourth}
In this section we give a brief summary
of our paper. We formulated the pure
spinor BRST charge redefinition in the
classical manner in order to be able to
generalize this to the case of pure
spinor string in general background.
Then we developed the Hamiltonian
formalism for pure spinor string in
general background and we found
Virasoro constraints. Then we analyzed
the general structure of the Poisson
brackets and discussed conditions under
which the classical redefinition of the
BRST charge can be performed.

The motivation for this calculation was
to see how Virasoro constraints can
emerge from the pure spinor string in
general background. In fact, it seems
to be rather difficult to study the
classical equations of motion for pure
spinor string without imposing Virasoro
constraints.
 The reason why we are interested
in the study of classical
solutions of pure spinor string
is following.
It is well known that the
 classical description of the
Green-Schwarz superstring in
$AdS_5\times S^5$ \footnote{For review,
see
\cite{Tseytlin:2003ii,Plefka:2005bk}.}
gives very interesting results and
predictions. Then it would be certainly
very interesting to give a covariant
form of this analysis using pure spinor
formulation of superstring. Then due to
the lack of quantum mechanical
formulation of pure spinor conformal
field theory in this background we
wanted to perform classical analysis of
pure spinor string in $AdS_5\times S^5$
as well. However it turned out that the
fact that Virasoro constraints are
"hidden" in the original pure spinor
formulation makes the classical
analysis rather obscure. On the other
hand we hope that the formulation of
the pure spinor theory in $AdS_5\times
S^5$ based on new BRST operator $Q'$
could be useful for description of the
classical dynamics of the pure spinor
string. We currently study this problem
and we hope to report about new results
in future.

\section{Appendix: Classical Poisson brackets}
In this Appendix we collect some
classical Poisson brackets between
fundamental modes for pure spinor
string in flat background.
To
begin with we define graded Poisson
bracket. Let as consider extended phase
space that is spanned with canonical
pairs $X^M,\Pi_M$ with Grassman parity
$|M|$. Then the
 graded Poisson bracket
 is defined as
\begin{equation}\label{gpbs}
\pb{F,G}= (-1)^{|F||M|}
\left[\frac{\partial^L F}{\partial X^M}
\frac{\partial^L G}{\partial \Pi_M}
-(-1)^{|M|}\frac{\partial^L F}{\partial
\Pi_M} \frac{\partial^L G} {\partial
X^M}\right] \ ,
\end{equation}
where superscript $L$ on partial
derivative means partial left
derivative and where the relation
between left and right derivative can
be found as follows. Let $F$ is
function of Grassmann parity $|F|$
defined on superspace labeled with
$X^M$.  Since
$dF(Z)=dX^M\partial^L_MF=\partial^R_MF
dX^M$ we obtain that left and right
derivatives of $F$ are related as $
(-1)^{|M||M+F|}\partial^L_M F
=\partial^R_M F$. In what follows we
will consider the derivative from the
left only and for that reason we omit
the superscript $L$ on the sign of the
partial derivative. Note also that the
Poisson brackets (\ref{gpbs}) obey
relation
\begin{eqnarray}\label{pbfg}
\pb{F,G}= -(-1)^{|F||G|}\pb{G,F} \
\nonumber \\
\end{eqnarray}
and generalized Jacobi  identity
\begin{equation}\label{Jacobi}
\pb{M,\pb{N,P}}+(-1)^{|M||N|+|M||P|}
\pb{N,\pb{P,M}}+(-1)^{|N||P|+|N||M|}
\pb{P,\pb{M,N}}=0 \ .
\end{equation}
Now we return to the pure spinor string
in flat background. Using the canonical
Poisson brackets (\ref{cpb}) we easily
determine the BRST variations of
fundamental modes
\begin{eqnarray}
\pb{Q_L,x^m}&=& i(\lambda
\gamma^m\theta) \ , \quad
\pb{Q_L,p_m}=-i\partial_\sigma (\lambda
\gamma^n\theta)
\eta_{mn} \ ,  \nonumber \\
\pb{Q_L,\theta^\alpha}&=& -\lambda^\alpha
\ , \quad  \pb{Q_L,\lambda^\alpha}=0 \ , \quad
\pb{Q_L,\pi_\alpha}=d_{\alpha} \   \nonumber \\
\end{eqnarray}
while Poisson bracket between $Q_R$ and
unheated variables all vanish. The same
relations can be derived in case of
$Q_R$ and hatted variables.

Further, we easily determine
\begin{eqnarray}
\pb{d_\alpha(\sigma),d_{\beta}(\sigma')}&=&
2i\gamma^m_{\alpha\beta}
(p_m-\partial_\sigma x^n\eta_{nm}
+2i(\theta\gamma^n\partial_\sigma\theta)\eta_{nm})
\delta(\sigma-\sigma')\equiv \nonumber
\\
&\equiv&
2i\gamma^m_{\alpha\beta}\Pi_{m-}\delta(\sigma-\sigma')
\ ,
\nonumber \\
\pb{\hd_{\halpha}(\sigma), \hd_{\hbeta}
(\sigma')}&=&
 2i\gamma_{\halpha\hbeta}^m
(p_m+\partial_\sigma x^n\eta_{nm} -2i
(\htheta \gamma^n\partial_\sigma
\htheta)\eta_{nm})\delta(\sigma-\sigma')\equiv
\nonumber \\
&\equiv&  2i\gamma_{\halpha\hbeta}^m
\Pi_{+m}\delta(\sigma-\sigma') \ ,
\nonumber \\
\pb{\hd_{\halpha}(\sigma),
d_{\beta}(\sigma')}&=&0 \ .
\end{eqnarray}
 Further, we have
\begin{eqnarray}
\pb{d_\alpha(\sigma),\Pi_{m-}(\sigma')}&=&
-4i(\gamma^n\partial_\sigma\theta)_\alpha
\eta_{nm}\delta(\sigma-\sigma')
\nonumber \\
\pb{\hd_{\halpha}(\sigma), \Pi_{m+}
(\sigma')}&=& 4i(\gamma^n\partial_\sigma
\htheta)_{\halpha}\eta_{nm}\delta(\sigma-\sigma')
\nonumber \\
\pb{d_{\alpha}(\sigma),\theta^\beta(\sigma')}&=&
-\delta_\alpha^\beta
\delta(\sigma-\sigma')  \ , \nonumber \\
\pb{\hd_{\halpha}(\sigma),
\htheta^{\halpha}(\sigma')}&=&
-\delta_{\halpha}^{\hbeta}\delta(\sigma-\sigma')
\nonumber \\
\pb{\Pi_{-m}(\sigma),\Pi_{-n}(\sigma')}&=&
-2\partial_\sigma
\delta(\sigma-\sigma')\eta_{nm} \ , \nonumber \\
\pb{\Pi_{+m}(\sigma),\Pi_{+n}(\sigma')}&=&
2\partial_\sigma
\delta(\sigma-\sigma')\eta_{mn}
\nonumber \\
\pb{\Pi_{-m}(\sigma),\Pi_{n+}(\sigma')}&=&0
\nonumber \\
\end{eqnarray}
With the help of these Poisson brackets
we obtain
\begin{eqnarray}
\pb{Q_L,Q_L}= 2i\int d\sigma
\lambda^\alpha\lambda^\beta
\gamma^m_{\alpha\beta}\Pi_{m-} \ ,
\nonumber \\
\pb{Q_R,Q_R}= 2i \int d\sigma
\hlambda^{\halpha}\hlambda^{\hbeta}
\gamma_{\halpha\hbeta}^m \Pi_{+m} \ .
\nonumber \\
\end{eqnarray}
Further, using the Poisson brackets
given above we easily get
\begin{eqnarray}
\pb{Q,\Pi^m_-}&=& -4i(\lambda
\gamma^m\partial_\sigma\theta) \ , \quad
\pb{Q,\Pi^m_+}=
4i(\hlambda\gamma^m\partial_\sigma\htheta)
\nonumber \\
\pb{Q,d_\alpha}&=&
2i(\lambda\gamma^m)_\alpha \Pi_{m-} \ , \quad
\pb{Q,\hd_{\halpha}} =2i
(\hlambda\gamma^m)_{\halpha}\Pi_{m+} \
.
\nonumber \\
\end{eqnarray}
Now we determine the Poisson
brackets between $Q$ and ghost
variables $N, J$ where
\begin{eqnarray}
N_{mn}= \frac{1}{2}\pi_\alpha
(\gamma_{mn})^\alpha_\beta\lambda^\beta
\ , \quad J=\pi_\alpha\lambda^\alpha \
, \nonumber \\
\hat{N}_{mn}= \frac{1}{2}\hpi_{\halpha}
(\gamma_{mn})^{\halpha}_{\hbeta}
\hlambda^{\hbeta} \ , \quad
\hJ=\hpi_{\halpha}\hlambda^{\halpha} \
. \nonumber \\
\end{eqnarray}
Using the free Poisson brackets defined
above we easily get
\begin{eqnarray}
\pb{Q,N_{mn}}=\frac{1}{2} (\lambda
\gamma_{mn}d) \ , \pb{Q,J}= d\lambda \
, \nonumber \\
\pb{Q,\hN_{mn}}=\frac{1}{2} (\hlambda
\gamma_{mn}\hd) \ , \pb{Q,J}=
\hd\hlambda \ . \nonumber \\
\end{eqnarray}
Finally it is also useful to know
the Poisson bracket between $N_{mn}$'s
\begin{eqnarray}
\pb{N_{mn}(\sigma),N_{kl}(\sigma')}
= (\eta_{kn}N_{ml}-N_{mk}
\eta_{nl}-\eta_{mk}N_{nl}
+\eta_{lm}N_{nk})
\delta(\sigma-\sigma') \ .  \nonumber \\
\end{eqnarray}
\\
\\
{\bf Acknowledgement}

This work
 was supported  by the Czech Ministry of
Education under Contract No. MSM
0021622409.

\newpage

\end{document}